*Article*

# Capacity, Collision Avoidance and Shopping Rate under a Social Distancing Regime

Haitian Zhong [1] and David Sankoff [2],*

1 University of Ottawa; hzhong3@uottawa.ca
2 University of Ottawa; sankoff@uottawa.ca
* Correspondence: sankoff@uottawa.ca

**Abstract:** Capacity restrictions in stores, maintained by mechanisms like spacing customer intake, became familiar features of retailing in the time of the pandemic. Shopping rates in a crowded store under a social distance regime is prone to considerable slowdown. Inspired by the random particle collision concepts of statistical mechanics, we introduce a dynamical model of the evolution of shopping rate as a function of a given customer intake rate. The slowdown of each individual customer is incorporated as an additive term to a baseline value shopping time, proportional to the number of other customers in the store. We determine analytically and by simulation the trajectory of the model as it approaches a Little's Law equilibrium, and identify the point beyond which equilibrium cannot be achieved. By relating customer shopping rate to the slowdown compared to the baseline, we can calculate the optimal intake rate leading to maximum equilibrium spending. This turns out to be the maximum rate compatible with equilibrium. The slowdown due to the largest possible number of shoppers is more than compensated for by the increased volume of shopping. This macroscopic model is validated by simulation experiments in which avoidance interactions between pairs of shoppers are responsible for shopping delays.

## 1. Introduction

Economic recovery during the COVID-19 pandemic led to the imposition of safety measures such as social distancing in indoor gathering places such as stores. Widely applied criteria included capacity limits, such as percent reduction of normal zoning or fire regulation capacities, such as 30% or 50% of normal. Other rules consisted of density restrictions, the number of persons per unit area, such as one person per 5 or 10 square metres. While there is sensible safety-based reasoning behind these limits, and much analytical and simulation work on relating store capacity (e.g., Chang et al [1], Ntounis et al [11]) or customer density within a store (e.g., Echeverría-Huarte et al [2], Harweg, Bachmann and Weichert [3], Mayr and Köster [8]) to physical distance restrictions, there is little theoretical work addressing the detailed relation between capacity or density restrictions and their financial impact on the enterprise. It is may be implicitly assumed that restricting the number of customers will simply have a proportional effect on the total amount of shopping that its done. We will argue, however, that this cannot be universally true.

In queuing theory and operations research, a rule, "Little's Law", which is claimed to hold universally, states that the average number of customers in a stationary system is equal to the intake rate times the average time a customer spends in the system – Little [6]. The key term here is "stationary". This rule does not apply directly to the startup period before stationarity is achieved, but neither does it take into account conditions that preclude stationarity, such as when the system capacity is exceeded. The limited applicability of the rule is widely understood: "The only requirements are that the system be stable and non-preemptive; this rules out transition states such as initial startup or shutdown [which means that the entire system and its components are in stable operation and are not affected by external events]. ... because a store in reality generally has a limited amount of space, it can eventually become unstable... if the arrival rate is much greater than the exit rate, the store will eventually start to overflow." – Wikipedia [12].

The effect of intake rate, the flow of customers into the store, on total shopping rate is the phenomenon that we study in this paper. We approach it in two very different ways. First, we introduce a simple dynamical model of the evolution of shopping rate as a function of a given customer intake rate, starting with an empty store. The slowdown of each individual customer is incorporated as an additive term to a baseline value shopping time, proportional to the number of other customers in the store. This is a minimal model linking "microscopic" behaviour to macroscopic quantities, reminiscent of statistical mechanics, with no variability in customer behaviour and no queuing issues. The only variables are the intake rate, the baseline shopping time, and a single proportionality coefficient that incorporates, without specifying any details, store area and layout, social distancing regime, and customer interaction (collision avoidance) effects. We determine analytically the trajectory of the model as it approaches a Little's Law equilibrium, and identify the point of phase change, where equilibrium cannot be achieved. By relating customer shopping rate to the slowdown compared to the baseline, We can calculate the optimal intake rate leading to maximum equilibrium spending, and whether this optimal rate is inside the domain of Little's Law equilibrium or is on the border of non-equilibrium behaviour of the model.

For the second approach, we add an actual store layout to the model as well as a stochastic effect attached to individual customers. We also include a set of randomized strategies, common to all customers, for avoiding infringements of physical distance requirements. It is this that determines the link between the microscopic and macroscopic aspects of the simulation, which was assumed in the purely analytical models.

As with our deterministic model, the only control exercised by the store is the timing of customer entrances. Increasing the entrance rate will increase the number of customers in the store, but it will also slow each customer down as they avoid breaching the social distance criterion. Given these countervailing effects, the enterprise wants to maximize total shopping volume.

In a simulation, each shopper "chooses" a random assortment of locations in the store and tries to complete her shopping efficiently while avoiding other customers. Each potential encounter requires a diversion or backtracking of one or both shoppers, slowing them down. This slowdown gets worse as the number of shoppers in the store increases, leading to instances of local gridlock (also termed jamming), until a major proportion of the shoppers will be frozen in place, unable to move without infringing on the space of other shoppers. The entrance rate where this occurs is analogous to the highest rate consistent with a Little's Law equilibrium in our macroscopic model.

There is a large scientific literature on pedestrian dynamics and simulation (e.g., Muramatsun and Nagatani [9], Kouskoulisa, Spyropouloua and Antoniou [5], Nagatani, Ichinose and Tainaka [10]), including studies by simulation of the effect of increasing the number of individuals moving in one direction, opposing directions or two intersecting directions, until the point of "traffic jamming". This extends to open source: Kleinmeier et al [4], Lopez et al [7] and commercial (e.g., PedSim, AnyLogic, Simwell, Oasys, Bentley Systems, PTV Group,...) applications of pedestrian simulation. None of this work, however, is directly relevant to our project for two main reasons. First, our focus, embodied in our objective function, namely the total volume of shopping during opening hours, does not seem to have a counterpart anywhere in the pedestrian simulation literature and software, where the interest is basically the smooth movement of large numbers of people. Second, in the crowd movement literature, the movement is either along corridors or walkways from point-to-point or the efficient entrance and exit of large number of people through bottlenecks or from an enclosed space, not the independent, multi-directional movement of shoppers, each of whom arrives with, or evolves, a complex, unpredictable, but coherent trajectory, in our study.

## 2. The model and phase change

Our macroscopic model is deterministic, involving the controlled rate of flow $f$ of customers through the enterprise and the fixed physical characteristics of the floor plan expressed by the effective area parameter $c$. Our only assumption is embodied in a simple equation expressing the average delay in shopping time, caused either by collision or, in our current application, by social distancing.

### 2.1. The model

Let $f = 1/\Delta$ be the flow rate, i.e., the number of shoppers entering per unit time (in minutes$^{-1}$). This is the only quantity controlled by the store manager.

Let $L$ be the length of shopping list (number of items) and $P$ the average price per item, in \$, so that $M = LP$ is the total expenditure by each customer.

We require a coefficient $c$ relating to store capacity to the frequency of encounters invoking social distance rules.

Let the variable $n$ be the number of shoppers in store at any particular time and the variable $A$ to refer to the store time, in minutes, for a customer. We fix $A_1$ to be the store time for a customer in an otherwise empty store, i.e., with no interactions with other customers.

Our model rests on a single assumption, that the rate of increase of shopping time over $A_1$ is due to the rate of encounters between shoppers, which is proportional to $\binom{n}{2}$, where $n$ is the number of shoppers in the store.

### 2.2. The equilibrium phase

Equilibrium holds if the number of shoppers entering a store is equal to the number exiting. We assume that the number of encounters for a given shopper is proportional to the (constant) number of other shoppers in the store, times the length of time she is in the store.

**Theorem 1.** *Equilibrium holds iff* $f \leq \frac{(c+1)^2}{4cA_1}$.

*If* $f = \frac{(c+1)^2}{4cA_1}$, *i.e. at the boundary of the domain of equilibrium,*

$$A = \frac{2c}{c+1}A_1 \quad (1)$$

$$n = \frac{c+1}{2} \quad (2)$$

**Proof.** Because of the equilibrium condition $n = fA$, the number of other shoppers entering while a given customer is in the store. This is an instance of Little's law. By definition, $A_1$ is the time it would take a shopper to complete her shopping were there no other shoppers in the store. Then, by assumption written before the Theorem 1,

$$\begin{aligned} A &= A_1 + \frac{(n-1)A}{c} \\ &= A_1 + \frac{fA^2 - A}{c} \end{aligned} \quad (3)$$

So:

$$\begin{aligned} cA &= cA_1 + fA^2 - A \\ 0 &= fA^2 - (c+1)A + cA_1 \\ A &= \frac{(c+1) \pm \sqrt{(c+1)^2 - 4fcA_1}}{2f} \end{aligned} \quad (4)$$

Thus, for equilibrium to hold,

$$f \leq \frac{(c+1)^2}{4cA_1}. \tag{5}$$

At equality, we obtain

$$f = \frac{(c+1)^2}{4cA_1}, \tag{6}$$

$$A = \frac{2c}{c+1}A_1, \tag{7}$$

and,

$$n = \frac{c+1}{2}. \tag{8}$$

□

Let $e$ = expenditure rate per customer = $M/A$, and $E = en$ =Total expenditure rate per unit time.

To maximize $E$ with respect to $\Delta$, given $M$ and $c$

$$E = nM/A \tag{9}$$
$$= fM \tag{10}$$

The equilibrium state that maximizes $E$ is then one where $f = \frac{(c+1)^2}{4cA_1}$ and $A = \frac{2c}{c+1}A_1$. However long the average shopper would take in an empty store, almost doubling this time would maximize expenditure rate under equilibrium conditions.

In the next two sections, we explore how this maximizing equilibrium, as well other equilibria that do not maximize $E$, are approached as customers enter the store (initially empty), one by one.

The fundamental assumption is this work, that the number of interactions between shoppers is proportional to the square of $n$, the number of shoppers, does not lead to analytical expressions for the number of shoppers in the store at any point of time, the time taken for shoppers to complete their visit to the store, or other quantities of interest. These may be calculated empirically for any particular setting of the parameters $M, f, A_1$ and $c$, but here we seek more general mathematical expressions for these quantities.

The difficulties arise from the fact that the number of customers, and hence the rate of interactions are step functions, with the steps at integer times. While this is tractable as long as no customers are finished shopping, as soon as the first few customers finish, this will generally occur at non-integer times. Then we start accumulating nested expressions containing "ceiling" and "floor" functions, having no simple form.

Our first approach in Section 3 is to allow the number of interactions to be proportional to $t^2$, where $i < t < i + 1$. This modification leads to a general solution to the problem, although this is expressed partly through recurrences.

The second approach is to retain the stepwise increase in the number of interactions, but to round the customers' shopping times to the nearest integer. This also leads to a general solution, although the limiting behaviour is oversimplified.

## 3. A continuous model

Starting with an empty store, we number the clients $i = 1, 2, \ldots$ according to the time $(i-1)/f$ they enter, which, in our deterministic model, is the same order in which they leave the store. The number of clients in the store at the time individual $j$ enters is $[j - K_j]$, where $K_j$ is the last individual to leave the store before $j$ enters. Prior to the time $J_1$ when the first shopper completes her shopping, $j - K_j = j$.

The spending rate for any individual in the store in the instant after individual $j$ enters, is $\frac{M}{A_1+[j-1-K_j]/c}$, so the total spending rate is $[j - K_j]\frac{M}{A_1+[j-1-K_j]/c}$.

For tractability, we introduce the approximation

$$\int_{x=0}^{J_1-1} \frac{1}{1+\frac{x}{cA_1}} dx = Mf, \tag{11}$$

in which the shopping rate deceases slightly in the interval between the entry of the $j - 1$-st and $j$-th shopper, instead of remaining constant. Then we immediately derive

$$Mf = cA_1 \log\left(1 + \frac{J_1-1}{cA_1}\right), \tag{12}$$

Solving this,

$$J_1 = cA_1(e^{\frac{Mf}{cA_1}} - 1) + 1 \tag{13}$$

While the first customer shops during $J_1 - 1$ intervals, the second customer enters when $x = 2$ and exits when $x = J_2$, shopping during $J_2 - 2$ intervals. Continuing with our approximate model,

$$Mf = \int_{x=1}^{J_1-1} \frac{1}{1+\frac{x}{cA_1}} dx + \int_{y=J_1-2}^{J_2-2} \frac{1}{1+\frac{y}{cA_1}} dy \tag{14}$$

$$= cA_1 \left[ \log(1 + \frac{J_1-1}{cA_1}) - \log(1 + \frac{1}{cA_1}) + \log(1 + \frac{J_2-2}{cA_1}) - \log(1 + \frac{J_1-2}{cA_1}) \right] \tag{15}$$

$$= cA_1 \log\left[\frac{(1+\frac{J_1-1}{cA_1})(1+\frac{J_2-2}{cA_1})}{(1+\frac{1}{cA_1})(1+\frac{J_1-2}{cA_1})}\right]. \tag{16}$$

From (12), we then have

$$1 + \frac{J_2-2}{cA_1} = (1 + \frac{1}{cA_1})(1 + \frac{J_1-2}{cA_1}) \tag{17}$$

$$J_2 = J_1 + \frac{cA_1 + J_1 - 2}{cA_1}. \tag{18}$$

The third customer enters when $x = 3$ and exits when $x = J_3$, shopping during $J_3 - 3$ intervals.

$$Mf = \int_{x=2}^{J_1-1} \frac{1}{1+\frac{x}{cA_1}} dx + \int_{y=J_1-2}^{J_2-2} \frac{1}{1+\frac{y}{cA_1}} dy + \int_{z=J_2-3}^{J_3-3} \frac{1}{1+\frac{z}{cA_1}} dz \tag{19}$$

$$= cA_1 \log\left[\frac{(1+\frac{J_1-1}{cA_1})(1+\frac{J_2-2}{cA_1})(1+\frac{J_3-3}{cA_1})}{(1+\frac{2}{cA_1})(1+\frac{J_1-2}{cA_1})(1+\frac{J_2-3}{cA_1})}\right] \tag{20}$$

We know from (12) and (17) that

$$\frac{1+\frac{J_2-2}{cA_1}}{1+\frac{J_1-2}{cA_1}} = 1 + \frac{1}{cA_1}, \tag{21}$$

so that

$$\frac{(1+\frac{1}{cA_1})(1+\frac{J_3-3}{cA_1})}{(1+\frac{2}{cA_1})(1+\frac{J_2-3}{cA_1})} = 1. \tag{22}$$

$$\frac{cA_1 - 3 + J_2}{cA_1 + 1} + J_2 = J_3 \tag{23}$$

Continuing in this way, we arrive at the recurrence:

**Theorem 2.** *For $r \leq J_1 - 1$*

$$\frac{cA_1 - r + J_{r-1}}{cA_1 + r - 2} + J_{r-1} = J_r \tag{24}$$

In general, a customer may leave the store at a non-integer time $J^*$, where $r - 1 < J^* < r$, for some $r \geq 1$. The shopping rate during the time interval $(r - 1, r)$ changes discontinuously at this point. To account for this, we extend the recurrence in Theorem 2 as follows.

**Theorem 3.** *For any $r > 1$, suppose $r - 1 < J^* < r$. Then:*

$$\frac{1}{cA_1 + J^* - 2 - K_{r-1}} = \frac{J_r - J_{r-1}}{cA_1 + J_{r-1} - r} \tag{25}$$

*For all $r > 1$, if there is no $J^*$ satisfying $r - 1 < J^* < r$, then:*

$$\frac{1}{cA_1 + r - 2 - K_{r-1}} = \frac{J_r - J_{r-1}}{cA_1 + J_{r-1} - r} \tag{26}$$

**Proof.** Because the total shopping expenditure $M$ for all customers is the same, the spending for customers $r - 1$ in the schema below and $r$ is the same during the time period when both of them are in the store (B in the schema below, so the shopping expenditure for customer $r - 1$ from time $r - 1$ to time $r$ (A + B) equals to the shopping expenditure for customer $r$ from time $J_{r-1}$ to $J_r$ (B + C). So the spending in A equals the spendng in C.

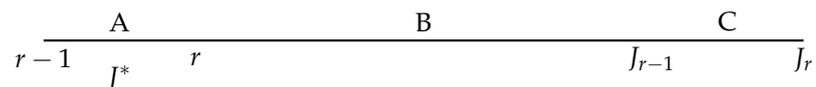

There are two cases. For all $r > 1$, if there exists a $J^*$ satisfying $r - 1 < J^* < r$, then:

$$\int_{x=r-2-K_{r-1}}^{J^*-1-K_{r-1}} \frac{1}{1+\frac{x}{cA_1}} dx + \int_{x=J^*-2-K_{r-1}}^{r-2-K_{r-1}} \frac{1}{1+\frac{x}{cA_1}} dx = \int_{y=J_{r-1}-r}^{J_r-r} \frac{1}{1+\frac{y}{cA_1}} dy$$

$$cA_1[\log\big(\frac{1+\frac{J^*-1-K_{r-1}}{cA_1}}{1+\frac{r-2-K_{r-1}}{cA_1}}\big)] + cA_1[\log\big(\frac{1+\frac{r-2-K_{r-1}}{cA_1}}{1+\frac{J^*-2-K_{r-1}}{cA_1}}\big)] = cA_1[\log\big(\frac{1+\frac{J_r-r}{cA_1}}{1+\frac{J_{r-1}-r}{cA_1}}\big)]$$

$$\frac{1}{cA_1 + J^* - 2 - K_{r-1}} = \frac{J_r - J_{r-1}}{cA_1 + J_{r-1} - r}$$

In the second case, for all $r > 1$, if there is no $J^*$ satisfying $r - 1 < J^* < r$, then:

$$\int_{x=r-2-K_{r-1}}^{r-1-K_{r-1}} \frac{1}{1 + \frac{x}{cA_1}} dx = \int_{y=J_{r-1}-r}^{J_r-r} \frac{1}{1 + \frac{y}{cA_1}} dy$$

$$cA_1 [\log \left( \frac{1 + \frac{r-1-K_{r-1}}{cA_1}}{1 + \frac{r-2-K_{r-1}}{cA_1}} \right)] = cA_1 [\log \left( \frac{1 + \frac{J_r-r}{cA_1}}{1 + \frac{J_{r-1}-r}{cA_1}} \right)]$$

$$\frac{1}{cA_1 + r - 2 - K_{r-1}} = \frac{J_r - J_{r-1}}{cA_1 + J_{r-1} - r} \qquad (27)$$

□

Theorems 2 and 3 permit us to calculate an exact description of the process given any values of the store and shopping parameters. These are not closed forms, but they allow a rapid and precise calculation of the numbers of shoppers at all times.

We can graphically display the behaviour of $J_r$ and related quantities using a specific example.

The parameter values should satisfy the following:

i) The parameter $A_1$ represents the total shopping time for a customer in an empty store, i.e., no other customers. So $\Delta$ should be less than $A_1$, considerably so for the model to be interesting. And $f = 1/\Delta$ represents the number of customers who enter the store per unit time. Thus:

$$A_1 * f > 1 \qquad (28)$$

ii) Because there will be a continual stream of customers entering the store, the shopping time for first customer, $\frac{J_1-1}{f}$, should be greater than $A_1$, so:

$$J_1 - 1 = cA_1(e^{\frac{Mf}{cA_1}} - 1) > A_1 * f$$

$$\implies M > \frac{cA_1}{f} * \ln \left( \frac{f}{c} + 1 \right) \qquad (29)$$

One realistic scenario is $M = 0.137, c = 70.58, A_1 = 0.2108$ (in hours). Changing the parameter values will change the scale of the results but not their basic form.

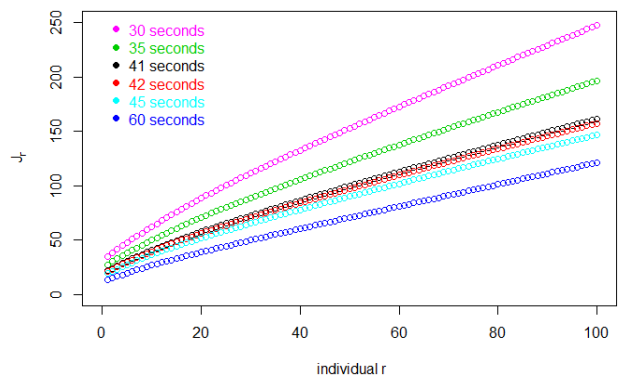

**Figure 1.** The value of $J_r$ for the first 100 customers.

Figure 1 shows $J_r$, the total number of customers entering before individual $r$ leaves the store. Then $\Delta(J_r - r)$ is the total shopping time for the $r^{th}$ customer. In this figure, for each $\Delta$, later customers, i.e., larger values of $r$, use increasing amounts of time to complete their shopping, because more customers are in the store, slowing down each customer's shopping rate. By the same token, as $\Delta$ decreases, the store fills up at a faster rate and the entire trajectory of $J_r$ takes on higher values. Figure 2 represents the number of customers in the store during the first 20,000 seconds, for various values of $\Delta$. In this figure, we find that each curve for each $\Delta \geq 42$ seconds asymptotically approaches a constant. For $\Delta \leq 41$ seconds, on the other hand, $J_r$ increases without bound. While the curves for all $\Delta > 42$ seconds tend to be infinite, which means the store will be more and more crowded ends up freezing up in the real life.

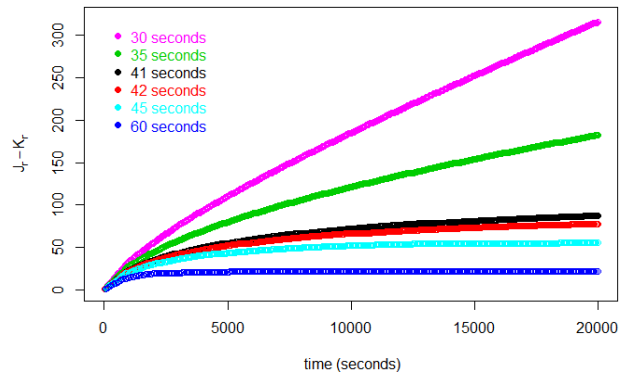

**Figure 2.** Number of customers in the store during 20000 seconds

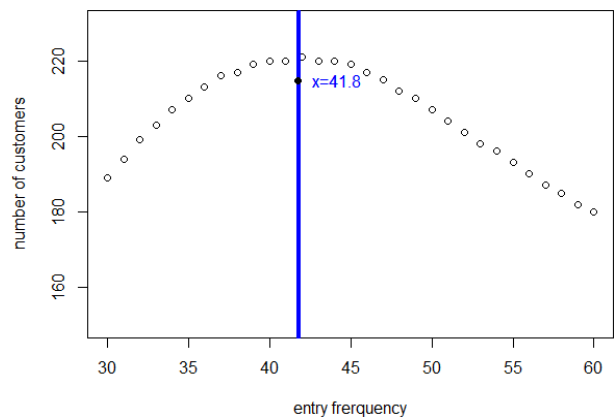

**Figure 3.** Number of customers who have completed shopping during 5000 seconds

In Figure 3, we could see the number of customers who have completed shopping during 5000 seconds. Although we did not define the judgment criteria for freezing in our mathematical model. We can draw some conclusions about this part from this figure. According to our assumption that customers will complete their shopping and leave the store one by one with the order that they entered the store. So for a store that operates normally without congestion, the higher flow rate, the more customers will complete the shopping during the same time interval. However, it's obvious that this curve reaches its maximum at 42 seconds. This reflects from the side that when flow rate is too large, it will cause serious congestion in the store which leads to the freezing, and when flow rate is small, the store can run stably which will tends to be equilibrium.

But why we get the maximum at 42 seconds and why we draw a line at 41.8 second, this will be explained in Section 4

On the other hand, the basic shape and the trend of the curve in Figure 4 will be seen to be similar to Figure 12 for a discrete model in the next section. But the absolute value of the this plot in the continuous model is somewhat greater than in the discrete model. This means customers need to use somewhat more time to complete shopping.

As expected, the absolute value of the average shopping time for each customer for the continuous model is higher than what we will find for the discrete model within the same entry interval. In Figure 5, those points points exceeding $A = \frac{2c}{c+1} A_1 = 1496.36$ seconds and the dividing point between more than $\geq A = \frac{2c}{c+1} A_1$ and $\leq A = \frac{2c}{c+1} A_1$ is exactly the boundary between equilibrium and non-equilibrium, just as we will see in the discrete case. Of course, that point is still 41.8 seconds which is calculated by the theory in Section 2. And the actual average shopping time is approaches $A = \frac{2c}{c+1} A_1$ when the entry interval is close to 41.8 seconds.

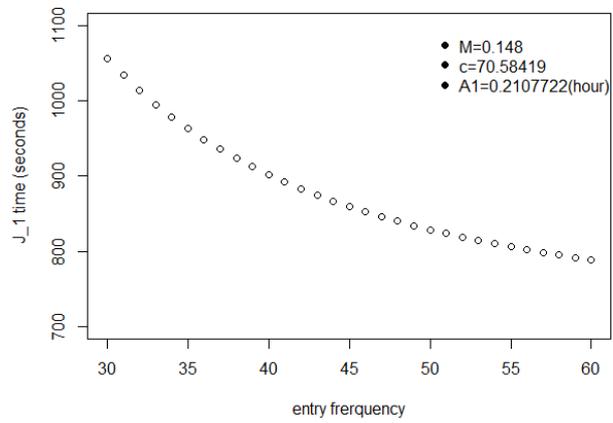

**Figure 4.** $J_1$ as a function of entrance interval.

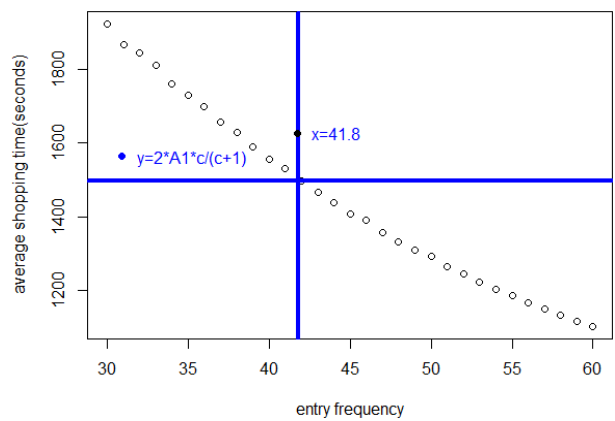

**Figure 5.** Average shopping time of customers who have completed shopping for different entry frequency

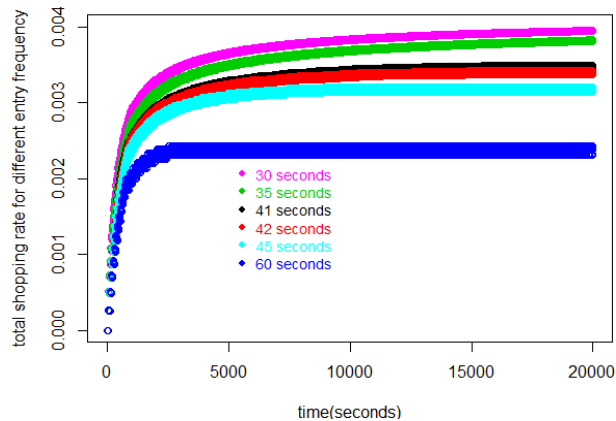

**Figure 6.** Total spending rate for all customers in the store at each time for various entry intervals, within 20000 seconds.

Looking at the rates for the $\Delta \geq 42$ curves in Figure 6. the basic trends of these curves are similar to the discrete model. But the absolute value of the line for the same entry interval in continuous situation is larger than in the discrete situation.

The overall conclusion is that the way to optimize total spending while keeping the number of shoppers stable, is to allow as many shoppers at to enter as possible, getting close as to the bifurcation point but not exceeding it.

## 4. The dynamics of store time

In this section, we explore an alternative formulation of the same problem discussed in Section 3, where the number of individuals in the store increases as a step function. instead of continuous growth. It is closer to reality with the assumption that the $J_r$ are all integer-values, compared to the non-integer-values in the continuous model.

*4.1. The initial trajectory*

The spending rate for the first individual after the $j^{th}$ individual enters is $\frac{M}{A_1+[j-1]/c}$ so that she buys at only $\frac{A_1}{A_1+\frac{j-1}{c}}$ of her best rate. Using this rate multiply by the length of time interval $\frac{1}{f}$ will be the expenditure during this time interval. Then she will have to wait until the $J_1$-st individual enters to finish spending $M$, where

$$M = \frac{1}{f} \sum_{j=1}^{J_1-1} \frac{A_1}{A_1 + \frac{j-1}{c}}$$

$$Mf = \sum_{j=1}^{J_1-1} \frac{A_1}{A_1 + \frac{j-1}{c}}$$

$$= \sum_{j=1}^{J_1-1} \frac{1}{1 + \frac{j-1}{cA_1}}$$

$$= \left( \sum_{k=1}^{J_1-2} \frac{1}{1 + \frac{k}{cA_1}} \right) + 1, \qquad (30)$$

under the assumption that $J_1$ is an integer. The effects of this assumption, which extends to the shopping completion times $J_2, \ldots$ of subsequent customers, are minimal, as long as $A_1$ is large with respect to $\Delta$, but they nonetheless motivate the continuous version of my model in Section 3. From 30, we have

$$\frac{Mf}{cA_1} = \psi(cA_1 + J_1 - 1) - \psi(CA_1), \tag{31}$$

where $\psi$ is the digamma function. As $x$ gets large $\psi(x)$ approaches $\log x$ so that $Mf$ is approximated by

$$cA_1 \log(1 + \frac{J_1 - 1}{cA_1}). \tag{32}$$

which is identical to the form (12) we derived in the continuous situation, representing the limiting behaviour of the discrete model. But we have found that this assumption represents a very slight departure from reality.

We now study the case of shoppers who enter the store before the first shopper leaves, namely shoppers $r = 2, \ldots, J_1 - 1$.

Examining when the second shopper leaves the store, i.e., when the $J_2$-nd shopper enters, we have

$$\begin{aligned}
Mf &= \sum_{j=2}^{J_1-1} \frac{1}{1 + \frac{j-1}{cA_1}} + \sum_{j=J_1}^{J_2-1} \frac{1}{1 + \frac{j-2}{cA_1}} \\
&= \sum_{k=1}^{J_1-2} \frac{1}{1 + \frac{k}{cA_1}} + \sum_{k=J_1-2}^{J_2-3} \frac{1}{1 + \frac{k}{cA_1}} \\
&= \sum_{k=1}^{J_2-3} \frac{1}{1 + \frac{k}{cA_1}} + \frac{1}{1 + \frac{J_1-2}{cA_1}}
\end{aligned} \tag{33}$$

From (30) and (33),

$$\begin{aligned}
\sum_{k=1}^{J_1-2} \frac{1}{1 + \frac{k}{cA_1}} + 1 &= \sum_{k=1}^{J_2-3} \frac{1}{1 + \frac{k}{cA_1}} + \frac{1}{1 + \frac{J_1-2}{cA_1}} \\
1 &= \sum_{k=J_1-1}^{J_2-3} \frac{1}{1 + \frac{k}{cA_1}} + \frac{1}{1 + \frac{J_1-2}{cA_1}} \\
&= \sum_{k=J_1-2}^{J_2-3} \frac{1}{1 + \frac{k}{cA_1}}
\end{aligned} \tag{34}$$

Examining when the third shopper leaves the store, i.e., when the $J_3$-rd shopper enters, we have

$$\begin{aligned}
Mf &= \sum_{j=3}^{J_1-1} \frac{1}{1 + \frac{j-1}{cA_1}} + \sum_{j=J_1}^{J_2-1} \frac{1}{1 + \frac{j-2}{cA_1}} + \sum_{j=J_2}^{J_3-1} \frac{1}{1 + \frac{j-3}{cA_1}} \\
&= \sum_{k=2}^{J_1-2} \frac{1}{1 + \frac{k}{cA_1}} + \sum_{k=J_1-2}^{J_2-3} \frac{1}{1 + \frac{k}{cA_1}} + \sum_{k=J_2-3}^{J_3-4} \frac{1}{1 + \frac{k}{cA_1}} \\
&= \sum_{k=2}^{J_3-4} \frac{1}{1 + \frac{k}{cA_1}} + \frac{1}{1 + \frac{J_1-2}{cA_1}} + \frac{1}{1 + \frac{J_2-3}{cA_1}}
\end{aligned} \tag{35}$$

From (33) and (35),

$$\sum_{k=1}^{J_2-3} \frac{1}{1+\frac{k}{cA_1}} + \frac{1}{1+\frac{J_1-2}{cA_1}} = \sum_{k=2}^{J_3-4} \frac{1}{1+\frac{k}{cA_1}} + \frac{1}{1+\frac{J_1-2}{cA_1}} + \frac{1}{1+\frac{J_2-3}{cA_1}}$$

$$\frac{1}{1+\frac{1}{cA_1}} = \sum_{J_2-3}^{J_3-4} \frac{1}{1+\frac{k}{cA_1}} \qquad (36)$$

The calculations above illustrate an induction argument by which we can prove:

**Theorem 4.** *For $r \leq J_1 - 1$,*

$$\frac{1}{1+\frac{r-2}{cA_1}} = \sum_{J_{r-1}-r}^{J_r-r-1} \frac{1}{1+\frac{k}{cA_1}} \qquad (37)$$

This is the analog of Theorem 2 in the continuous model.

*4.2. Approach to equilibrium*

Theorem 4 opens a way to the computation of the $J_r$ for the customers who enter before $J_1$. The situation is somewhat more difficult for the customers who enter later. We use the same idea used to prove Theorem 3.

We denote by $K_r$ the number of customers who have left the store up to and including the moment the $r$-th customer enters. So if $J_{k^*} \leq r < J_{k^*+1}$, then $K_r = k^*$.

Recalling our discussion of $K_r$ at the beginning of this Section, We rewrite Theorem 4 as

**Theorem 5.**

$$\frac{1}{1+\frac{r-(K_{r-1}+2)}{cA_1}} = \sum_{i=J_{r-1}-r}^{J_r-r-1} \frac{1}{1+\frac{i}{cA_1}} \qquad (38)$$

This version of the theorem allows us to compute $J_r$ (and $K_r$) for all $r$, not only those covered by Theorem 4. The recurrence in the theorem is not, however, a closed form that would allow us to find a value of $r$ for which equilibrium is attained.

Nevertheless, we can prove some strong results about $J_r$ and $K_r$ in the equilibrium phase. Since customers enter once every time interval, one customer has to leave every time interval. Then $J_r - J_{r-1} = 1$ if the equilibrium condition is met for $r - 1$ or earlier. With some effort we can prove:

**Theorem 6.** *If $J_r - J_{r-1} = 1$ occurs, then equilibrium is reached from time $J_{r-1}$ and the number of customers in the store will always satisfy $J_r - r$.*

*And $J_r = 2r - (K_{r-1} + 1)$ will always hold starting from some time $r_0$ which is later than time $J_1$.*

$J_r$ will eventually reach $J_r^{equilibrium} = 2r - (K_{r-1} + 1)$ and remain at that value. If $J_r$ does not satisfy this equation, equilibrium has not yet been reached.

To discover the exact condition for equilibrium, we study the difference between the value of $J_r$ calculated by Theorem 5 and the hypothetical value of $J_r$ were equilibrium reached at the outset. If this difference is 0 at some point, this means that $K_{r-1} = K_{r-2} + 1$ has been satisfied, and $J_r - J_{r-1} = 1$ will hold thenceforth.

With a considerable amount of work, we can prove:

**Theorem 7.** *If $g(r) = J_r - J_r^{equilibrium}$ starts to decrease at time $r_0$, which is earlier than time $J_1$, then it will reach equilibrium at time $J_{r_0}$ and the number of customers in the store will remain* $J_{r_0} - r_0$. *On the other hand, if $g(r)$ starts to increase before time $J_1$, then it will never reach equilibrium and the number of customers will always increase.*

We examine an example where $M = 0.148, c = 70.58, A_1 = 0.2108$ (in hours):

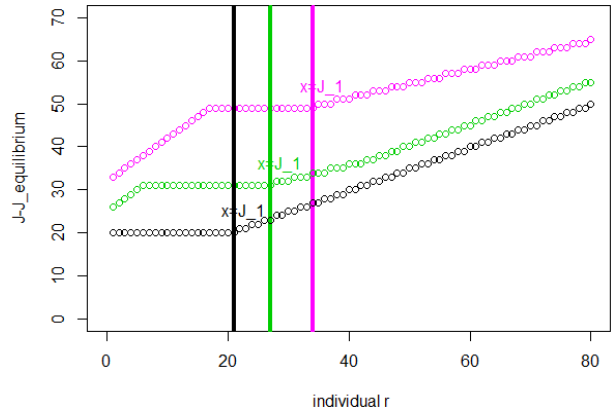

**Figure 7.** The difference between actual J and theoretical value of J at equilibrium for time intervals which is too big to be equilibrium

This picture (7) shows the value of $g(r)$ of 3 different entry intervals which could not achieve equilibrium. And the vertical lines of the corresponding colors represent $g(J_1)$. We find the value of $g(r)$ for 30 seconds and 35 seconds goes up first and then is flat for a while. But after time $J_1$ continuously increases although its upward trend has slowed. And for the 41 second interval, it is flat first and then goes up after time $J_1$.

Because we will use a rounding rule to choose an integer $J_r$ for the discrete situation in the next section, so there is some discrepancy, but the total trend of these 3 lines is consistent with what we will find there.

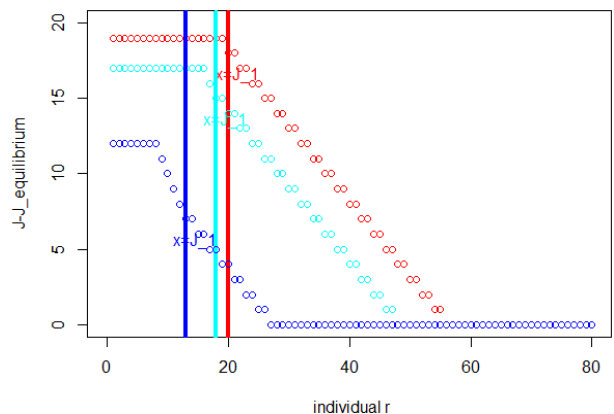

**Figure 8.** The difference between actual J and theoretical value of J at equilibrium for time intervals which is small enough to be equilibrium

In Figure (8), we drawn the line according to the value of $g(r)$ of 3 different entry intervals which lead to equilibrium. They are 42 seconds, 45 seconds and 60 seconds from top to bottom. We could find at time $J_1$, they all show a downward trend and they all remain 0 after a specific time. These are also consistent with what we have expect.

We can show

**Theorem 8.** *The total shopping rate will keep increasing until the maximum is achieved when equilibrium has been reached, although the shopping rate for individual customers will decrease.*

It is remarkable, as Theorem 6 shows, that if there is to be equilibrium when customer $r$ leaves, this is already pre-ordained by the dynamics of the number of customers ($K_{r-1} = K_{r-2} + 1$) in the store immediately before the $r$-th customer even enters.

Our characterization of the equilibrium context in Theorems 4, 5 and 6 can be identified with the parameters of Subsection 2.1:

$$A = \Delta(J_r - r) \tag{39}$$
$$n = J_r - r = 2r - (K_{r-1} + 1) - r = r - (K_{r-1} + 1) = r - K_r \tag{40}$$

To illustrate the effects of entry frequency on $J_1$, We picked particular values for the parameters $M, c$ and $A_1$, namely $M = 0.148, c = 70.58, A_1 = 0.2108$ (in hours). Though we could have picked a wide range of other values without changing the general form of the output variables, these particular numbers are in a range that is both realistic and amenable, in terms of computing time, to the simulations reported in Sections 5 and 6.

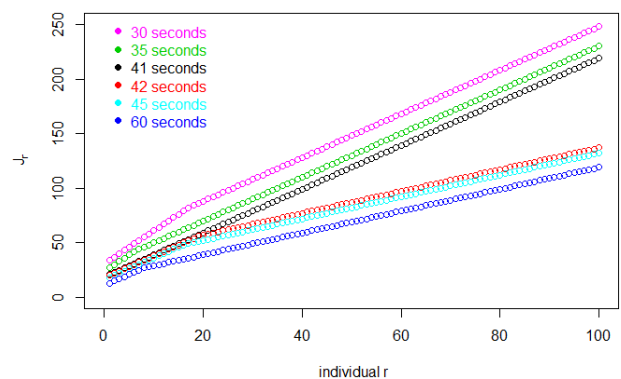

**Figure 9.** The value of $J_r$ for the discrete model

Recall that the $r^{th}$ customer enters the store at the beginning of the $r^{th}$ time interval. And the value of $J_r$ represents the number of time intervals which have passed when $r$ leaves. Multiplying $r$ or $J_r$ by $\Delta$ converts the "number of intervals" scale to clock time.

In working out my example, we had to come to terms with the fact that my formal development in this Section requires that the $J_r$ are integer-valued, which leads to small errors. For example the sum in (30) will exceed $Mf$, but if We left off the $J_1^{st}$ term, the sum would be smaller than $Mf$. My practical solution was to choose whichever sum was closest to $Mf$. In actual calculations, normalizing the interval between the two sums to $[0, 1]$, it turned out that the keeping the $J_1^{st}$ was preferable only when the sum was within 0.465 (not 0.5) of the desired value, and otherwise it was preferable to drop this term..

As with the continuous model, the value of $J_r$ will increase when the entry interval decreases. However we could see a big difference in figure 9.

This graph reveals a bifurcation in the pattern of the $J_r$ curves as $\Delta$ changes. The curves for all $\Delta \geq 42$ seconds follow a common pattern different from the common pattern for $\Delta \leq 41$ seconds. The slopes of the curves for the larger values of $\Delta$ tend to 1, while the slopes of the curves for the smaller values of $\Delta$ tend to 2. A slope greater than than 1 means customers are entering faster than earlier customers are leaving. This will result in continuously increasing numbers of customers in the store and the impossibility of equilibrium. This is clear in Figure 10, where the number $L_r$ of customers in the store when $r$ enters is:

$$L_r = r - K_r. \tag{41}$$

In figure 10, when $\Delta \geq 42$ seconds, $L_r$ eventually attains a Little's law equilibrium. But for $\Delta \leq 41$ seconds, $L_r$ increases without limit. My model does not require specifying store capacity, but this could be easily incorporated into our calculations to find the point at which the store will overflow, or customers will no longer be able to move around. In Section 5 of this thesis, We will show how this "pathology" plays out in a more realistic model.

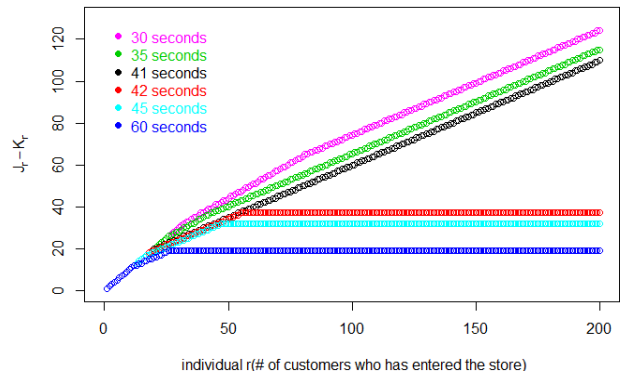

**Figure 10.** The number of customers in the store when individual *r* enters

In Subsection 2.2, We obtained conditions for equilibrium. From Theorem 1 and the particular model parameters used illustrate the model it can be calculated that the condition for equilibrium is

$$\Delta = 1/f \tag{42}$$

$$\geq \frac{4cA_1}{(c+1)^2} \tag{43}$$

$$= 41.8 \text{ seconds}, \tag{44}$$

which agrees with the bifurcation observed in Figures 9 and 10.

It is of instructive to plot the number of customers who complete shopping within a substantial time after opening, say 5000 seconds, as in Figure 11.

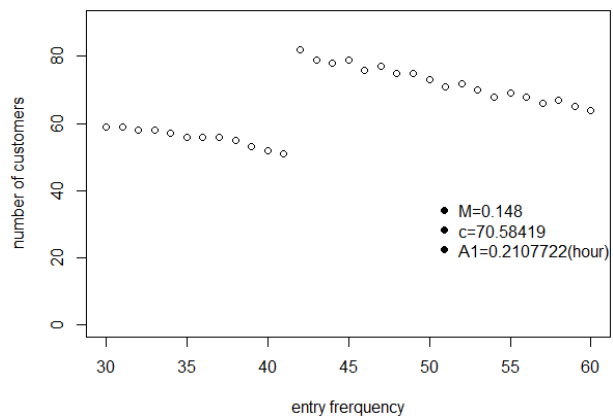

**Figure 11.** Number of customers who have completed shopping within 5000 seconds

When the entry interval $\Delta$ is small ($\leq 41$ seconds) or, equivalently, entry frequency $f$ is large, the number of customers who enter the store up to any given time may well be greater than with more severe restriction on entry, but the number of customers who succeed in completing their shopping is less when the enter interval is too small, namely 41 seconds or less in our model.

The number of customers who complete their shopping differs very much between $\leq 41$ seconds and $\geq 42$ seconds. This is because, if there is no equilibrium, the rapidly rising number of customers will slow each other down and their spending rate will decrease so fast that few of them can complete their total shopping list. As for entry interval which is $\geq 42$ seconds, customers complete their shopping lists quickly and at a steady rate. Thus, the number of customers who complete their shopping will be greater even though fewer customers will have entered the store.

This is why we got the maximum value of the number of customers who completed the shopping during 5000 seconds at 42 seconds in Figure 3. From here, we could find the discrete model is better than continuous model to describe this realistic problem.

Another interesting plot depicts $J_1$ with different entry intervals.

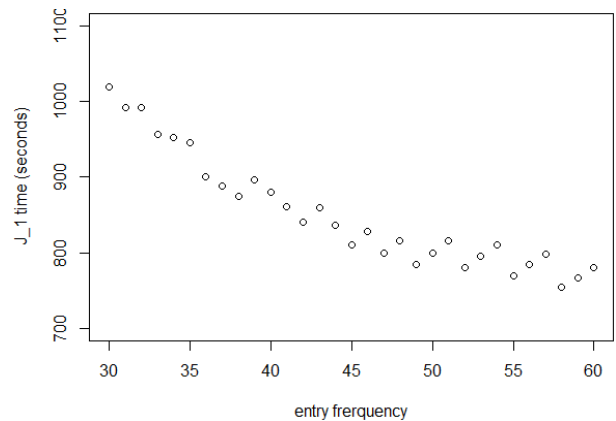

**Figure 12.** First customer exit time $J_1$.

The *x*-axis of Figure 12 is in clock time, measured in seconds. All the $J_1$ in this figure are larger than 759 seconds which is the same as $A_1$. This attests to the reasonableness of the selection of the value of the parameter values.

Note that $J_1$ is not the point where equilibrium is reached, but represents a slower rate of increase in shopping time for large $\Delta$, but not for small $\Delta$.

The shopping time for the customer after the first customer must be longer than the time for the first one. In the figure, the shopping time for the first customer with a 30-second entry interval is 250 seconds more than the one with a 60 entry interval. That is almost $\frac{1}{3}$ of the shopping time for first customer with a 60 second entry interval. This means that for the 30-second entry interval the customer has already been affected by a lot of other customers.

One important thing is that according to the Theorem 7, when we want to make sure if equilibrium could be reached with a specific $\delta$, we just need to pay attention to the customers who entered the store before $J_1 * \delta$. If there are any two of these customers, like $r_0$ and $r_0 + 1$ who leave the store 1 minute apart, then equilibrium could be reached at time $J_{r_0}$.

The saw-tooth effect visible in this figure is an artifact of our $J_1$ rounding, which also affects the next few figures; it has no interest beyond that.

Figure 13 shows a plot of the average shopping time of a customer who has completed shopping within 5000 seconds. This is parameter $A$ in Subsection 2.

Figure 13 includes a horizontal and a vertical line. These highlight the fact that when $\Delta \geq 42$ seconds, the average shopping time of customers who have completed shopping list will be$< \frac{2*A_1*c*3600}{c+1} = 1496.36$ seconds. That the maximum shopping time under equilibrium is $A = \frac{2*A_1*c}{c+1}$ can be understood in terms of Theorem 1. On the other hand, if the entry interval $\leq 41$ seconds, the average shopping time will be much greater.

Besides that, these data could help to compare formula (40) with $n = fA$ used in Theorem 1. The data in Figure 13 is $A$. A comparative picture for entry intervals that are $\geq 42$ seconds because they approach equilibrium.

In this figure, the black points are from the calculation of the discrete model, and this kind of data is obtained from Figure 10 directly. Then the red points are obtained from the product of the data plotted in Figure 13 and the corresponding entry interval. These two sets of data are very close. There is a little error between them because of the approximation when calculating $J_r$. But this does not affect the final conclusion, that is, our formula (40) and theory in Subsection 2.1 are consistent.

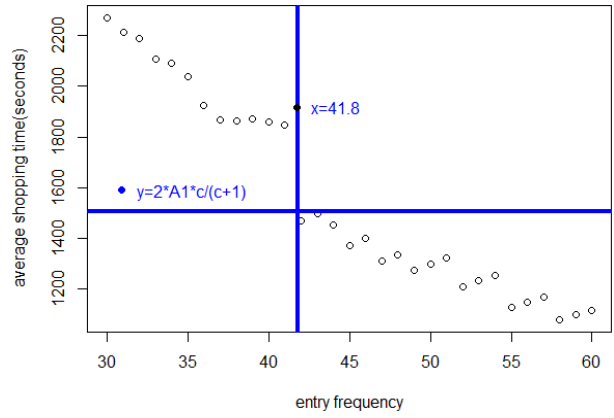

**Figure 13.** The average shopping time of customers who has completed the shopping for different entry frequencies.

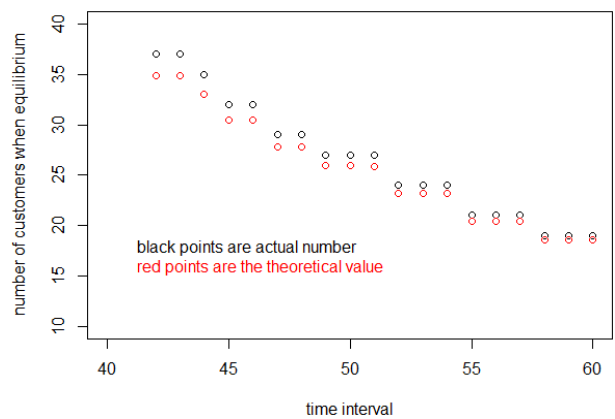

**Figure 14.** The number of customers in the store when it meets equilibrium.

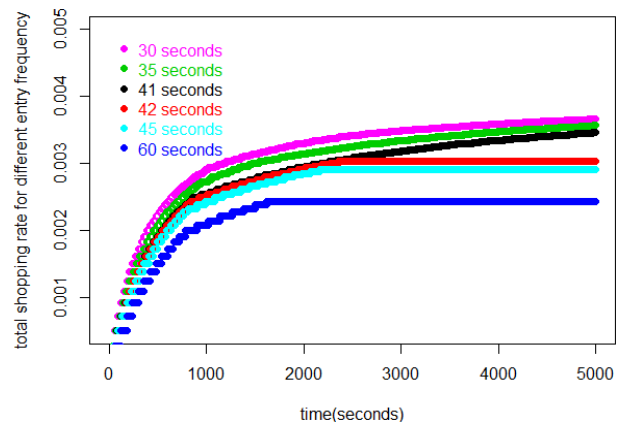

**Figure 15.** Total spending rate for the customers in the store for various entry intervals.

We conclude this section with the most important results, the total spending rate. This is depicted in Figure 15, which illustrates the sum of the spending rate of each customer currently in the store. Two patterns can be distinguished. For those ∆ which are too small to result in an equilibrium flow of customers, without equilibrium, their number will keep increasing unrealistically and there will be no stable spending rate.

The other pattern in Figure 15 includes the total spending rate when ∆ is from 42 seconds to 60 seconds. These curves will be flat after a specific time and this means they will all have stable spending rates. Moreover, the total spending rate will increase as the entry interval decreases. Although there will be more customers in the store which slows down the spending rate for each customer, this effect is not large enough to negate the positive effect of *n*, the stable number of the customers. Our total spending rate is equal to the number of customers multiplied by each customer's spending rate. With this model, the conclusion is that letting as many customers in the store as is compatible with stability will be the most profitable.

## 5. Design of the simulations
*5.1. Overview*

We model the freedom of shoppers to formulate their own shopping lists in advance, with a random, normally distributed, number *L* of items located at random points along the display cases around the store. This placement of the items also models a passive role of management. The customer enters and travels from one of their list items to the closest next one, using a line of sight approach, or proceeding to the end of the aisle (the closest corner of an internal display) until completion, when they exit. In the course of their shopping, customers may have to pause or deviate from their trajectory to avoid collisions (i.e. breaches of the social distance criterion). The dispersed distribution of *L* instead of the fixed total cost *M* in the discrete and continuous models is a change that no longer allows the assumption that customers enter and leave in the same order.

The store is modelled as a rectangular layout with one four-sided display in the centre. There would be little difficulty in incorporating layouts involving any number of displays, counters, walls, aisles (possibly one-way), checkout lines, other waiting lines, and even multiple floor levels, without substantially changing the objective function, the control parameter (entrance timing) or the simulation principles. In contrast, the discrete and continuous models needed no specific layout or even total area. All that was necessary were the parameters $A_1$ and *c*.

Included with the simulation is an animated visualization. Each customer is represented by a coloured dot, and her shopping points are represented by colour-coded ×'s along the boundary, as in Figure 16. The trajectory of the customers and their interactions can be followed in the actual time scale. The appearance and disappearance of shopping points are visible, and a separate display shows an alert each time there is an interaction, with the ID's of the two customers involved.

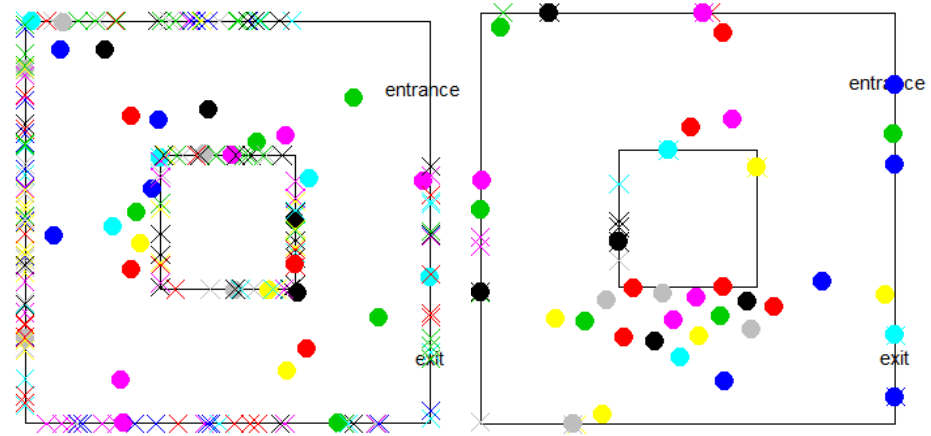

**Figure 16.** Frames from the animation showing the distribution of customers at shopping points and in the aisles. On the right is an example of freezing. There are 14 customers who cannot move. Note that there are few shopping points left.

The main difference between the models in Sections 2, 4 and 3 and the simulation discussed in the present section is that there is no longer the assumption that each of the $n$ customers has an equal chance, proportional to $\frac{1}{n-1}$, of an interaction with each other customer, so that interactions happen at a rate proportional to $\binom{n}{2}$. Instead, every customer pursues her own shopping goals in the store, based on an efficient trajectory from one item to another, randomly chosen at the outset. Whenever there is immediate danger of social distance infringement with another customer, one or both of them takes evasive action, which prolongs their shopping time.

Another important difference is how departures from stability are manifested. In the models, this simply shows up as an increase of the number of customers without limit. In the simulation it is seen as a freezing or "jamming" [9,10] situation in some part of the store so that a large number of shoppers are stuck and the simulation cannot be completed, as on the right in Figure 16. Indeed, much of the research on this project was dedicated to avoiding, or at least delaying, the freezing phenomenon, but given the strict infringement avoidance rules, together with the goal of each customer to get to a nearby shopping point, some freezing is inevitable. We believe that more sophisticated behavioural simulation and models of shopper variability will reduce or delay this phenomenon, but not by much.

Figure 16 shows a typical freezing situation on the right. Many customers are stuck around the lower part of the store. Some of them need to go out through exit - note that there are few shopping points left compared to the left panel, but some other customers are still shopping. Since they all have to maintain social distance, they find themselves unable to advance to their goals.

*5.2. Simulation details*

The set-up for the simulation is as follows:

1. We designed a 30 meters by 30 meters store with a 10 meters by 10 meters display counter in the centre. The shopping points could occur along all four store walls and all four edges of the display counter.
2. The number of items on the shopping list of each customer follows a normal distribution with a mean of 15 and a variance of 2. These items are spatially distributed uniformly and randomly along the walls and display case.

3. To avoid freezing around the entrance, each shopper's trajectory is initiated at one of her shopping points, chosen at random. Every customer will go to this shopping point first, then complete the original shopping list as described below. This is implemented primarily to avoid repeated premature termination of simulations, but it may also be interpreted as the customer's desire to avoid waiting for her first shopping point if there are too many customers.
4. After each shopping point, including the first one, the customer chooses the nearest line-of-sight shopping point remaining on the list, if there is one, from her unfinished shopping list to be the next shopping point. The customer will follow a straight line between her current location and next shopping point.
5. Customers' walking speed is set at 0.3 meters per second. If the customer approaches a shopping point and the distance between her and the shopping point is smaller than 0.3 meters, then her next step is set to be this shopping point. Customers spend 15 seconds at every shopping point.
6. If none of the remaining items on her list are in the line of sight, as they may be obscured by the display table, the customer walks in a straight line to the next corner of the display table, and then proceeds to the nearest visible shopping point.
7. The motivation for this work being the impact on store management of a social distancing regime during the COVID-19 pandemic, the maintenance of physical distance is of paramount importance. The simulation is designed so that customers cannot get within two meters of each other.
8. To maintain the two-meter social distance, the customers follow a number of rules to avoid infringing on each other's space. If a step were to move a customer to within two metres of another person, this step is not taken. Instead, the program consults the other person's projected trajectory in relation to the current customer. According to which of four quadrants that movement is projected to be, the current customer chooses right (R), left (L), back (B) or wait (W) with the given probabilities, as in Figure 17. If an R, L, B move were to cross a wall or a display edge, it is replaced by W. If the customer chooses B, a number $s$ of steps backwards will be randomly taken with probability of $1/2^s$.

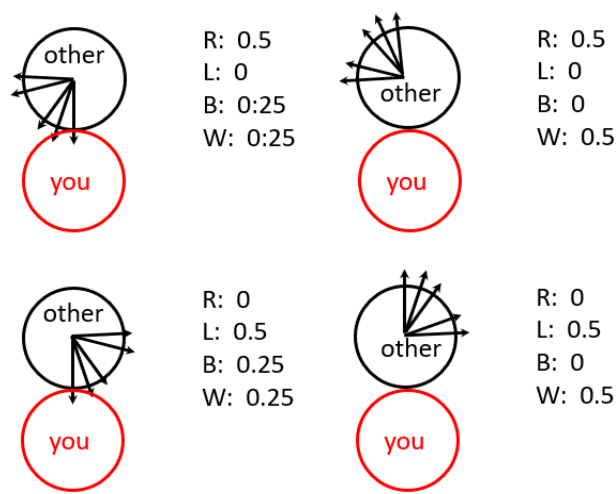

**Figure 17.** Rules for customers avoiding space infringement. The axis between the two shoppers at the projected point of infringement defines four quadrants, in which the "other" person is known to be be moving. The probabilities "you" chooses right (R), left (L), back (B) or wait (W) depends on the quadrant.

9. At each point of time, the movement of the customers is calculated sequentially, in the order they entered the store.

*5.3. Reducing the quadratic coefficient for collision detection*

Since the number of customers who can fit in the store is bounded, the computing time requirements for a simulation will only be proportional to the duration of the experiment. However, even with a moderate number of customers in the store, it is necessary to avoid calculating all pairwise distances between them too often, though for any two customers in the store at the same time, this distance must be calculated at least once. To keep to this minimum, We make use of a "risk time" linked list, which contains the shortest possible time that the distance between two customers can be reduced to meters.

Each customer $X$ carries two vectors. The first vector **risktime**$_X(Y)$ contains the collision risk time $t'_X(Y) = t'_Y(X)$. This number is usually calculated only once, when $X$ enters the store (if $Y$ is already there).

$$t'_X(Y) = \frac{\mathbf{dist}(X,Y)}{2 \times \mathbf{speed}} - \mathbf{social\ distance}. \tag{45}$$

Only when $t'_X(Y) = t$, the current time, is dist$(X, Y)$ calculated again, so that either we perform collision avoidance (rare) or update $t'_X(Y)$ (more usual). Only $t'_X(Y)$ is recalculated, not all the other customers. When $X$ leaves the store, all $t'_X(Y)$ are set to 1,000,000.

The second vector carried by $X$, called **nextrisk**$_X(Y)$ contains the name/number $Z$ which has the next risk time with $X$ immediately after $Y$. Finally there are two numbers associated with $X$, called $r(X)$ and $s(X)$, which contains the earliest time in $t'_X(Y)$ and the $Y$ which has this earliest time.

When $X$ enters the store, $t'_X(Y)$ is calculated for all $Y$. Then these values are ordered enabling us to define **nextrisk**$_X(Y)$ and $r(X)$. Also, for all the other $Y$ in the store, we have to find the largest $t'_Y(Z)$ less than $t'_X(Y)$, where **nextrisk**$_Y(Z) = W$ and change these to **nextrisk**$_Y(Z) = X$ and **nextrisk**$_Y(X) = W$.

Then $X$ starts shopping until time $r(x)$. At that time, we check $s(X)$. If $s(X) = Y$, we calculate dist$(X, Y)$. If this is greater than **social distance**, we recalculate $t'_X(Y)$ and insert it appropriately into **nextrisk**$(X)$. Otherwise we do collision avoidance before recalculating $t'_X(Y)$. Then we change $r(X)$ and $s(X)$ by consulting **nextrisk**$_X(Y)$. And $X$ continues shopping.

Note that any time we change $t'_X(Y)$ and adjust **nextrisk**$(X)$, we have to change $t'_Y(X)$ and adjust **nextrisk**$(Y)$.

Also note that when $X$ enters, all of the $t'_X(Y)$ must be calculated, and they must be ordered, to construct the linked list **nextrisk**$(X)$. But once $X$ starts shopping, no more long calculation. If $X$ is far from everybody else, $r(t)$ will be large and there will be many steps with no calculation. All we have to do is occasionally update $t'_X(Y)$ and $r(t)$ and maybe **nextrisk**$(X)$ and $s(t)$, which are just a few steps.

## 6. Results

For the simulations, 16 values of $\Delta$ were sampled, from 30 to 65 in five-second intervals, as well as 28 seconds, 31, 32, 33, 34, 37, 42 and 47. The unequal spacing was designed to span intervals where important changes in behaviour were expected. For each of these values, 100 successful simulation experiments were conducted, with customer positions recalculated every second up to 3600 seconds. "Successful" means that no freezing took place. All unsuccessful simulations, terminated by freezing events, were discarded and replaced by additional experiments.

To determine rates of freezing, in Figure 18, the proportion of all experiments, successful and unsuccessful, are plotted. Thus the 40% successful simulations where $\Delta = 28$ seconds is based on the 100 experiments to be used in subsequent analyses, plus 150 discarded unsuccessful experiments.

Operationally, if a customer stays at one point for more than 200 seconds, it is considered a freezing situation. If a customer enters the store at time $i$, but is unable to move from the entrance at time $i + \Delta$, this was also considered freezing.

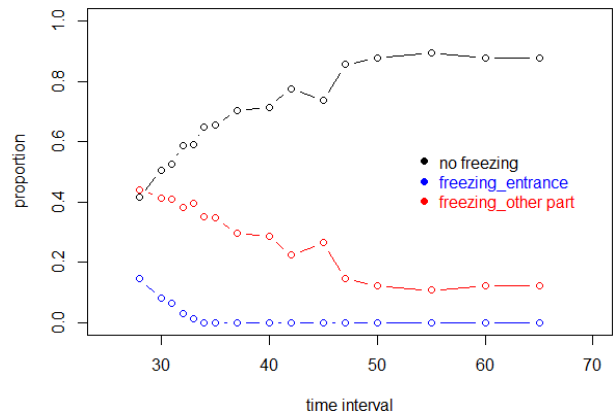

**Figure 18.** The rate of freezing as a function of entrance interval Δ.

The practice of discarding unsuccessful experiments is unavoidable. Replacing them until a successful experiment is recorded, is a choice made in the name of assuring enough data to accurately characterize the model. Both practices undoubtedly introduce biases into estimates, but how severe these are is not clear. We conjecture this gives results that portray the simulations with small Δ as being more favourable to total spending.

From Figure 18, it is clear that when entry interval decreases so that more customers enter the store per unit time, the probability of a successful experiment will decrease.

For all the variables we consider, we take the mean over 100 simulations for each of the 3600 seconds of the simulation.

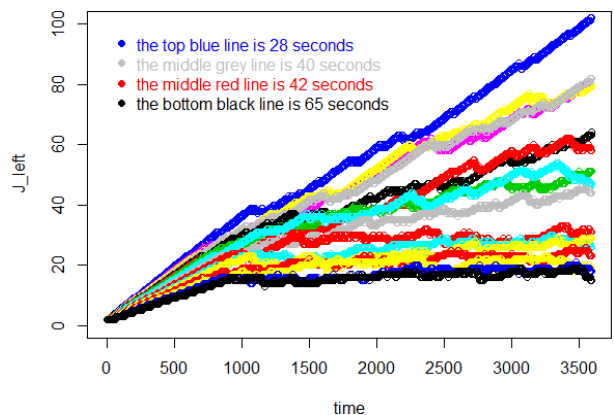

**Figure 19.** The number of customers in the store during 3600 seconds

In Figure 19, the line for Δ = 40 seconds is the grey line in the middle and the red line below it represents 42 seconds. A blank area is apparent between these two lines. Each of the curves below this blank area tend towards a constant value, while the other curves display increasing growth. This means that the number of customers in the store is stable after the initial growth when the entry interval is equal or bigger than 42 seconds. But for entry intervals which are equal or smaller than 40 seconds, although 3600 seconds worth of data can be extracted from these experiments, growth will eventually succumb to freezing. Already for 28 seconds, this happens more than half of the time.

Figure 20 reveals how many customers have completed their shopping list from time 0 to time 3600. This curve increases first and decreases later with the maximum value occurring in between. The decrease is understood from Figure 21, simply because there are fewer customers entering the store, with negligible decrease in shopping time. But why does the curve in Figure 20 increase at first? The reason for this situation could be found in Figure 19. That is because more and more customers stay in the store, the probability of collision for each customer increases a lot so they cannot access the shopping point directly.

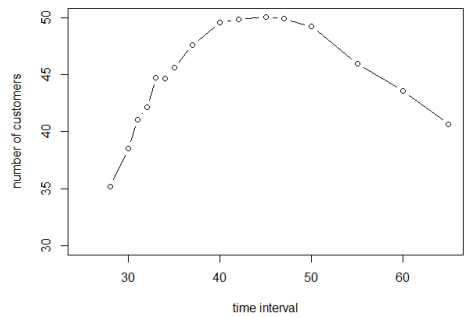

**Figure 20.** Number of customers who have completed shopping during 3600 seconds

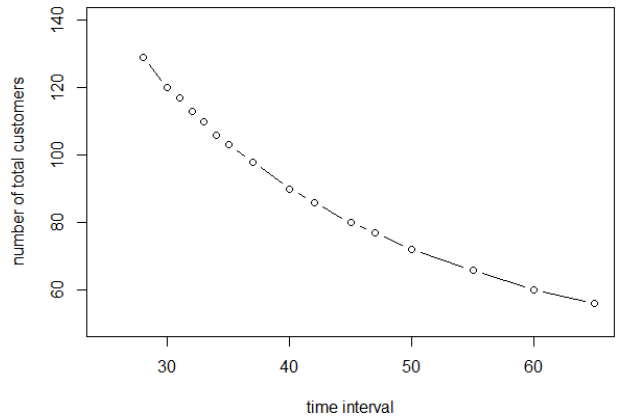

**Figure 21.** the number of customers who have entered the store within 3600 seconds

As a result, customers need more time to complete shopping list and fewer customers will complete before 3600 seconds.

In addition, the largest value is 50.04 which occurs when ∆ is 45 seconds. This means that an average of 50.04 people complete shopping during 3600 seconds when the time period was 45 seconds. However, the number of customers who completed shopping are all greater than 49 for all ∆ from 40 to 50 seconds. The effect of these entry intervals is similar.

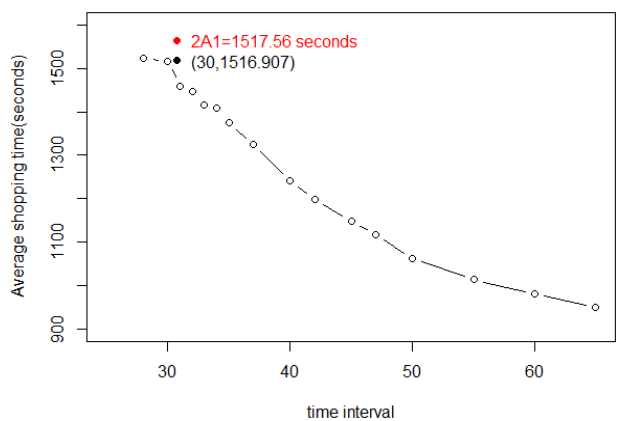

**Figure 22.** Average shopping time of customers who have completed shopping for different entry frequency

Figure 22 shows how long customers take to complete their shopping (only those who have completed their shopping before 3600 seconds). This curve keeps decreasing because when the entry interval is larger, the number of customers entering the store in the same time interval will decrease, and the probability of collision will decrease. In this way, customers arrive at their shopping points and exit without many detours. Their shopping time will approach $A_1 = 758.78$ seconds.

*6.1. Accessibility*

The simulation codes in this subsection can be found in the following link:
https://github.com/css614/Thesis-codes/blob/main/Thesis-code.R

These codes run in RStudio 3.6.1. The scale of the store and customers' walking speed are preset. So the most important parameter in these codes is the "entrance interval". When this decreases, it will take a longer time to run the whole code and it will be more susceptible to freezing. If an experiment freezes, an error code is generated.

Output generated and stored by this code includes the total number of customers who enter in the store, the location of each customer at every second, the shopping list of each customer, the shopping time of each customer, the spending rate of each customer at every second, the times of direction change by each customer and the state of each loop.

## 7. Conclusions

Whether or not our simulation experiments are realistic, given that customers behave identically except for their entrance time, and their shopping lists, is a subjective question. Certainly the animations give the impression of normal activity - even to the extent that the observer is tempted to impute strategies, personality traits and emotional states to some of the shoppers, none of which is included in the code!

More objectively, to what extent do the analytical models, lacking any tracking or analysis of customer movement, reflect the simulations, which model spatial dynamics in detail, including social distancing? This is basically a test of the fundamental assumption of the models that pairwise interactions causing slowdown will occur at the rate of $\binom{n}{2}$.

The most salient aspect of the models is their bifurcated character. This also appears in the simulation, and at the same value of $\Delta$ when $M$, $c$ and $A_1$ are the same. In the example, if $\Delta \geq 41.8$ seconds, there is an equilibrium state, but if $\Delta < 41.8$, there is no equilibrium. Moreover, in comparing simulation and theory, the number of customers at equilibrium for 60 seconds is around 20, for both approaches, and for 42 seconds is around 40.

The average shopping time for each customer having completed shopping decreases when the entry interval increases. Clearly, with fewer customers in the store, shopping is completed more quickly.

In both approaches the number of customers who complete shopping will increase first and decrease later as the entry interval $\Delta$ increases. And the largest number will occur at the bifurcation point.

The conclusion that emerges is that the "$\binom{n}{2}$-interactions" assumption suffices to produce the same bifurcated response to $\Delta$ as the more realistic simulation where the interactions are the result only of the random placement of shopping points together with social distancing rules.

Practically speaking, the consequences of this work is to provide an lower limit to the timing of customer entries. This value is the lowest that will permit a stable "Little's law" flow of customers. And our results shed light on the time course of the approach to equilibrium.

While the concepts incorporated in the models and the simulations have been illustrated with particular values of the parameters, the mathematics remain valid for all positive $M$, $c$ and $A_1$. There is not likely to be any closed form solution for $J_1$ and the other key quantities in our analysis. A hint of this lies in the connection with the digamma function.

**Funding:** This research was funded by an NSERC Discovery Grant to DS.

**Conflicts of Interest:** The authors declare no conflict of interest.

*Entropy* **2023**, *1*, 0          25 of 25